\pdfoutput=1
\documentclass{elsarticle}

\usepackage{amssymb}
\usepackage{amsthm}
\usepackage{amsmath}
\usepackage{epsf}
\usepackage{graphicx}
\usepackage{setspace}
\usepackage{color}
\usepackage{rotating}
\usepackage{longtable}
\usepackage[boxed, algoruled, nofillcomment, scleft, linesnumbered]{algorithm2e}

\theoremstyle{plain}

\theoremstyle{definition}

\begin{document}

\title{An improved Branch-and-cut code for the maximum balanced subgraph of a signed graph}

\author[ua]{Rosa Figueiredo\corref{cor1}}
\author[uff]{Yuri Frota}

\cortext[cor1]{Corresponding author. Fax number: +351~234370066 Email: rosa.figueiredo@ua.pt \\
Rosa Figueiredo is supported by FEDER founds through COMPETE-Operational Programme Factors of Competitiveness and by Portuguese founds
through the CIDMA (University of Aveiro) and FCT, within project PEst-C/MAT/UI4106/2011 with COMPETE number FCOMP-01-0124-FEDER-022690.
}

\address[ua]{CIDMA, Department of Mathematics, University of Aveiro \\ 3810-193 Aveiro, Portugal.\\
              \texttt{rosa.figueiredo@ua.pt}}
\address[uff]{Department of Computer Science, Fluminense Federal University \\ 24210-240 Niter\'oi-RJ, Brazil.\\
              \texttt{yuri@ic.uff.br}}

\begin{abstract}
The Maximum Balanced Subgraph Problem (MBSP) is the problem of finding a subgraph of a
signed graph that is balanced and maximizes the cardinality of its vertex set.
We are interested in the exact solution of the problem: an improved version of a branch-and-cut algorithm is proposed.
Extensive computational experiments are carried out on a set of instances from three applications previously discussed in the
literature as well as on a set of random instances.\\
\noindent\textbf{Keywords:} Balanced signed graph; Branch-and-cut; Portfolio analysis; Network matrix; Community structure.
\end{abstract}

\maketitle
\newpage

\section{Introduction}
\label{sec:introduction}


Let $G=(V,E)$ be an undirected graph where $V=\{1,2,\ldots,n\}$ is the set of vertices and
$E$ is the set of edges connecting pairs of vertices.
Consider a function $s:E\rightarrow\{+,-\}$ that assigns a sign to each edge in $E$.
An undirected graph $G$ together with a function $s$ is called a {\it signed graph}.
An edge $e\in E$ is called {\it negative} if $s(e)=-$ and {\it positive} if $s(e)=+$.

In the last decades, signed graphs have shown to be a very attractive
discrete structure for social network researchers~\cite{abell09,doreian96,doreian09,inohara98,Yang07}
and for researchers in other scientific areas, including portfolio analysis in risk management~\cite{Harary03,huffner09},
biological systems~\cite{DasGupta07,huffner09}, efficient document classification~\cite{bansal02},
detection of embedded matrix structures~\cite{gulpinarI} and community structure~\cite{Macon12,Traag09}.
The common element among all these applications is that all of them are defined in a collaborative vs.
conflicting environment represented over a signed graph.
We refer the reader to~\cite{zaslavsky98} for a bibliography of signed graphs.
Is this work we consider the Maximum balanced subgraph problem (MBSP) defined next.


Let $G=(V,E,s)$ denote a signed graph and
let $E^-$ and $E^+$ denote, respectively, the set of negative and positive edges in $G$.
Also, for a vertex set $S\subseteq V$, let $E[S]=\{(i,j)\in E\mid i,j\in S\}$ denote
the subset of edges induced by $S$.
A signed graph $G=(V,E,s)$ is {\it balanced} if its vertex set can be partitioned into
sets $W$ (possibly empty) and $V\setminus W$ in such a way that $E[W]\cup E[V\setminus W] = E^+$.
Given a signed graph $G=(V,E,s)$, the MBSP is the problem of finding a subgraph $H=(V',E',s)$ of $G$ such that
$H$ is balanced and maximizes the cardinality of $V'$.


The MBSP is known to be an NP-hard problem~\cite{barthold} although the problem of detecting
balance in signed graphs can be solved in polynomial time~\cite{harary80}.
In the literature, the MBSP has already been applied in the detection of embedded matrix
structures~\cite{Figueiredo12MBS,Figueiredo11,gulpinarI}, in portfolio analysis in risk management~\cite{Figueiredo12MBS} and community 
structure~\cite{Figueiredo12MBS}.

The problem of detecting a maximum embedded reflected network (DMERN) is reduced to the MBSP in~\cite{gulpinarI}.
Most of the existing solution approaches to the MBSP were in fact proposed for the solution of the DMERN problem.
The literature proposes various heuristics for the solution of the DMERN problem (for references see~\cite{gulpinarI}).
Lately, Figueiredo et al.~\cite{Figueiredo11} developed the first exact solution approach for the MBSP:
a branch-and-cut algorithm based on the signed graph reformulation from Gulpinar et al.~\cite{gulpinarI} for the DMERN problem.
Computational experiments were carried out over a set of instances found in the literature
as a test set for the DMERN problem. Almost all these instances were solved to optimality in a few seconds
showing that they were not appropriate for assessing the quality of a heuristic approach to the problem.
Recently, Figueiredo et al.~\cite{Figueiredo12MBS} introduced applications of the MBSP
in other two different research areas: portfolio analysis in risk management and community structure.
These authors also provided a new set of benchmark instances of the MBSP (including a set of difficult instances for the DMERN problem)
and contributed to the efficient solution of the problem by developing a pre-processing
routine, an efficient GRASP metaheuristic, and improved versions of a greedy
heuristic proposed in~\cite{gulpinarI}.

In this work we contribute to the efficient solution of the MBSP by developing
an improved version of the branch-and-cut algorithm proposed by Figueiredo et al.~\cite{Figueiredo11}.
We introduce a new branching rule to the problem based on the odd negative cycle inequalities.
Moreover, we improve the cut generation component of the branch-and-cut algorithm by
implementing new separation routines and by using a cut pool separation strategy. 

The remainder of the paper is structured as follows.
The integer programming formulation and the branch-and-cut algorithm proposed in~\cite{Figueiredo11} to the MBSP 
are outlined in Section~\ref{sec:ILP_BC}.
The improved version of the branch-and-cut algorithm is described in
Section~\ref{sec:BC_improved}.
In Section~\ref{sec:Computational}, computational results are reported for random instances as
well as for instances of the three applications previously mentioned.
In Section~\ref{sec:Conclusion} we present concluding remarks.

We next give some notations and definitions to be used throughout the paper.
For an edge set $B\subseteq E$, let $G[B]$ denote
the subgraph of $G$ induced by $B$.
A set $K\subseteq V$ is called a {\it clique} if each pair of vertices in $K$ is joined by an edge.
A set $I\subseteq V$ is called a {\it stable set} if no pair of vertices in $I$ is joined by an edge.
We represent a cycle by its vertex set $C\subseteq V$.
In this text, a signed graph is allowed to have parallel edges but no loops.
Also, we assume that parallel edges have always opposite signs.

\section{Integer programming formulation and branch-and-cut}
\label{sec:ILP_BC}

The integer programming formulation and the branch-and-cut algorithm introduced in~\cite{Figueiredo11} are described next.

\subsection{Integer programming formulation}

It is well known that a signed graph is balanced if and only if it does not contain a parallel edge or a cycle with an
odd number of negative edges~\cite{barahona89,gulpinarI,zaslavsky98}.
Let $C^o(E)$ be the set of all odd negative cycles in $G$, i.e., cycles with no parallel edges and with an odd
number of negative edges.
Throughout this text, a cycle $C\in C^o(E)$ is called an {\it odd negative cycle}.
The formulation uses binary decision variables $y\in\{0,1\}^{|V|}$ defined in the following way.
For all $i\in V$, $y_i$ is equal to 1 if vertex $i\in V$ belongs to the balanced subgraph, and is
equal to 0 otherwise.
We use the vector notation $y=(y_i)$, $i\in V$, and the notation $y(V')=\sum_{i\in V'} y_i$ for $V'\subseteq V$.
The formulation follows.
\begin{align}
\mathrm{Maximize\ }    & y(V)                      & \label{fo}\\
\mathrm{subject\ to\ } & y_i + y_j \leq 1, &\forall\ (i,j)\in E^-\cap E^+, \label{paralEdge}\\
                       & y(C) \leq |C|-1,  &\forall\ C\in C^o(E), \label{C-oddIneq}\\
                       & y_i\in\{0,1\},    &\forall\ i\in V.\label{yInt}
\end{align}
Consider a parallel edge $(i,j)\in E^-\cap E^+$. Constraints~(\ref{paralEdge})
ensure vertices $i$ and $j$ cannot belong together to the balanced subgraph.
Constraints~(\ref{C-oddIneq}), called {\it odd negative cycle inequalities}, forbid cycles with an odd number of negative
edges in the subgraph described by variables $y$. These constraints
force variables $y$ to define a balanced subgraph.
Finally, the objective function (\ref{fo}) looks for a maximum balanced subgraph.
The formulation has $n$ variables and, due to constraints~(\ref{C-oddIneq}), might have an exponential
number of constraints.
Let us refer to this formulation as $Y(G,s)$.
By changing the integrality constraints~(\ref{yInt}) in formulation $Y(G,s)$ by the set of trivial inequalities 
$0 \leq y_i \leq 1$, $i\in V$, we obtain a linear relaxation to the MBSP.

\subsection{A branch-and-cut algorithm}

The branch-and-cut algorithm developed in~\cite{Figueiredo11} is based on formulation $Y(G,s)$, 
uses a standard 0--1 branching rule and has three basic
components: the initial formulation, the cut generation and the primal heuristic.

\paragraph{\bf Initial formulation}

The initial formulation is defined as
\begin{align}
\mathrm{maximize\ }    & y(V)              &                 \nonumber\\
\mathrm{subject\ to\ } & y(K) \leq 1,      &\forall\ K\in L, \label{IF:1}\\
                       & y(C) \leq |C|-1,  &\forall\ C\in M\subseteq C^o(E), \label{IF:2}\\
                       & y(K) \leq 2,      &\forall\ K\in N, \label{IF:3}\\
                       & 0\leq y_i\leq 1,  &\forall\ i\in V, \label{IF:4}
\end{align}
where~(\ref{IF:1}) are clique inequalities from the stable set problem~\cite{Rebennack08} defined over a set of cliques $L$ in $G[E^+\cap E^-]$; 
(\ref{IF:2}) is a subset of inequalities~(\ref{C-oddIneq}) defined over a set of odd negative cycles $M$; 
(\ref{IF:3}) is a subset of inequalities from a family of negative clique inequalities introduced in~\cite{Figueiredo11} for the MBSP and defined
over a set of cliques $N$ in $G[E^-]$;
(\ref{IF:4}) is the set of trivial inequalities.
Greedy procedures described in~\cite{Figueiredo11} are used to generate sets $L$, $M$ and $N$.

\paragraph{\bf Cut generation}
After an LP has been solved in the branch-and-cut tree, the algorithm check if the solution is integer feasible. 
If this is not the case, the cut generation procedure is 
called and a set of separation routines is executed (a limit of 100 cuts per iteration is set). If no violated
inequality is found or if a limit of 10 cut generations rounds is reached, the algorithm enter in the branching phase.
The cut generation component described in~\cite{Figueiredo11} has two separation procedures.
An exact separation procedure is used to generate violated odd negative cycle inequalities~(\ref{C-oddIneq}).
This separation routine is based on a polynomial algorithm described in~\cite{barahona} to solve the separation
problem for cut inequalities.
A heuristic separation procedure defined in~\cite{Figueiredo11} is used to generate violated clique inequalities also introduced 
in~\cite{Figueiredo11}.

\paragraph{\bf Primal heuristic and branching rule}
A rounding primal heuristic is executed in~\cite{Figueiredo11} every time a fractional solution is found.
Moreover, a standard 0--1 branching rule is used with the same branching
priority assigned to each variable and the branch-and-cut tree is investigated with the best-bound-first strategy.
The authors reported they have also implemented a version of the branching
rule proposed in~\cite{balas86}. Although this branching rule has been successfully applied to
solve the stable set problem, they obtained better results with the standard 0--1 branching rule.

\section{An improved branch-and-cut code}
\label{sec:BC_improved}

In this work, the following new routines were added to the branch-and-cut algorithm described in Section~\ref{sec:ILP_BC}.

\paragraph{\bf Branching on the odd negative cycle inequalities}
Our branching rule is based on the odd negative cycle inequalities~(\ref{C-oddIneq}).
The intuition behind this cycle based branching is the attempt to generate more balanced enumerative
trees. The standard 0--1 branching rule can be very asymmetrical producing unbalanced enumerative trees. 

Let $\bar{y}\in\mathbb{R}$ be the optimal fractional solution of a node in the search tree.
Let $C'\subseteq C^o(E)$ be the subset of odd negative cycles such that each cycle $C\in C'$ satisfy the following conditions:

\begin{itemize}
\item constraint~(\ref{C-oddIneq}) defined by $C'$ is a binding one in the current formulation,
\item there exists a vertex $i\in C'$ such that $\bar{y}_i$ is fractional.
\end{itemize}

The standard 0--1 branching rule is used whenever $C'$ is an empty set.
If it is not the case, let $\bar{C}$ be the smallest cycle in $C'$.
Split $\bar{C}$ into the sets $\bar{C}^1$ and $\bar{C}^2$
such that $\bar{C} = \bar{C}^1\cup \bar{C}^2$, $\bar{C}^1\cap \bar{C}^2=\emptyset$ and $y(\bar{C}^1)$ is fractional.
We create three branches in the search tree:

\begin{itemize}
\item[(i)]  $y(\bar{C}^1) \leq  |\bar{C}^1|-1$ and $y(\bar{C}^2) = |\bar{C}^2|$;
\item[(ii)]  $y(\bar{C}^1) =  |\bar{C}^1|$ and $y(\bar{C}^2)\leq |\bar{C}^2|-1$;
\item[(iii)] $y(\bar{C}^1) \leq  |\bar{C}^1|-1$ and $y(\bar{C}^2)\leq |\bar{C}^2|-1$.
\end{itemize}

\paragraph{\bf Separation routines}
In this work, we introduce two new separation procedures to the cut generation component of the branch-and-cut algorithm
described in Section~\ref{sec:ILP_BC}.

The authors in~\cite{Figueiredo11} proved that lifted odd hole inequalities (from the stable set problem) defined over the set
of parallel edges $E^+\cap E^-$ are valid inequalities for the MBSP. They have also proved that, if the support
graph of these inequalities satisfy certain conditions they are facet defining inequalities to the problem.
We implemented a separation procedure described in~\cite{padberg73} to the lifted odd hole inequalities.
Also, the authors indicated in~\cite{Figueiredo11} that a very similar lifting procedure could be applied to 
strengthen constraints~(\ref{C-oddIneq}).
We implemented this lifting procedure to the odd negative cycle inequalities satisfying $|C|\leq 20$.
In both cases, a very small instance of the MBSP must be solved at each iteration of the lifting procedures. 
In our implementation, these small problems were solved by simple enumerative algorithms.

Moreover, we added a cut pool to the branch-and-cut code: any violated inequality included to the active formulation
of a node in the branch-and-cut tree is also included to the cut pool. 
As we have mentioned in Section~\ref{sec:ILP_BC}, after an LP has been solved in the branch-and-cut tree, we
check if the solution is integer feasible. If this is not the case, the cut generation procedure is then
called. Before running any separation routine from our cut generation procedure, we check if there are violated
cuts in the cut pool. In positive case, no separation routine is called and the violated cuts 
(limited to 100 cuts) are immediately added to the active formulation.

\section{Computational experiments}
\label{sec:Computational}

We implemented the improved branch-and-cut algorithm described in Section \ref{sec:BC_improved} 
using the formulation defined by (\ref{IF:1})-(\ref{IF:4}). Both branch-and-cuts (BC), the previous one and the improved version, were implemented in C++ running on a
Intel(R) Pentium(R) 4 CPU 3.06 GHz, equipped with 3 GB of RAM. We use
Xpress-Optimizer 20.00.21 to implement the components of these enumerative algorithms.
The maximum running time per instance was set at 3600 seconds. The same instance classes 
reported in \cite{Figueiredo12MBS} were tested here to allow for a better comparison of the performances of the improved BC and the BC algorithm proposed earlier. The class {\em Random} consists
of 216 randomized instances divided into two groups: Group 1 without parallel edges and Group 2 with parallel edges. The class {\em UNGA} is composed of 63
instances derived from the community structure of networks representing voting on resolutions 
in the United Nations General Assembly. The class new {\em DMERN} consists of 316 signed graphs coming from a set of general mixed integer programs. Finally, the class {\em Portifolio} is composed by 850 instances generated from market graphs. The entire benchmark is available for download in {\texttt{www.ic.uff.br/$\sim$yuri/mbsp.html}}.

We first investigate the behavior of the {\em Random} instances, the results obtained by the two methods are summarized in Table \ref{resRandomInstBC}. This table exhibits, for both groups, average times per $\mid V \mid$, and
percentage gaps per $\mid V \mid$, $d$ (density of the graph) and the rates $\mid E^- \mid / \mid E^+ \mid$ and 
$\mid E^+ \cap E^- \mid$. Multicolumn Time, gives us average times (in seconds) spent to solve instances
to optimality; the values in brackets show the number of instances solved to
optimality  (``-'' means no instance was solved within
the time limit). Multicolumn \%Gap presents the average of percentage gaps calculated 
over the set of unsolved instances. The percentage gap of each instance is
calculated between the best integer solution found and the final upper bound.
For each group of instances, the first and the second lines present, respectively,
the results obtained with the original and the improved code of the branch-and-cut algorithm. 
The results obtained with the improved version are slightly better: six
more instances were solved to optimality and all the average gaps were reduced.

\begin{figure}[Hp]
\hspace{-1cm}
\includegraphics[scale=0.5]{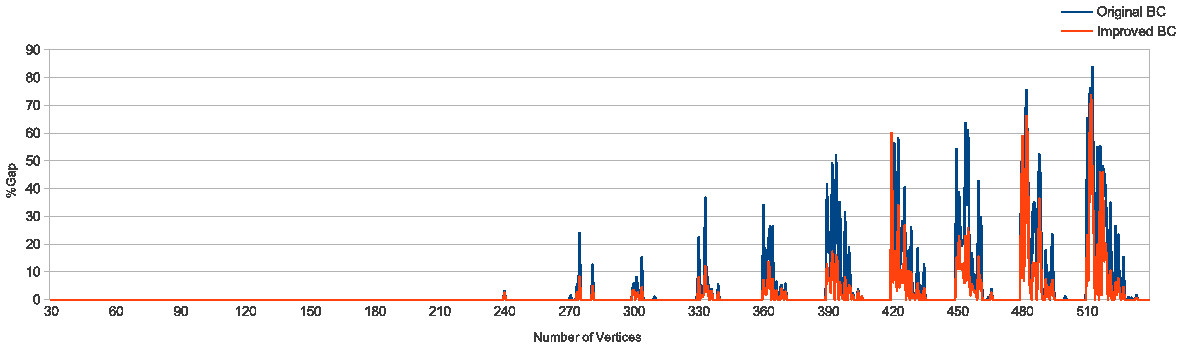}
\caption{\footnotesize Results obtained on portfolio instances.}
\label{resEconomyGraspxDFSxBFS}
\end{figure}

In the second experiment, we analyze the performance of the {\em Portifolio} instances. Table \ref{resBC_EconomyInstnaces} reports the obtained results. The first two columns give the number of vertices and a threshold value $t$ used to generate the instances \cite{Figueiredo12MBS}. The next three columns give the average time, the average of percentage gaps (as defined in Table \ref{resRandomInstBC}) and the number of evaluated nodes in the original BC tree, respectively. The last three columns give the same data for the improved BC.
Algorithm improved BC solved 227 out of 850 instances within
1 hour of processing time, while the original BC managed to solve only 217 instances. 
The average gap for the original BC over the set of unsolved instances is 17.91\%, while the same value for the improved version is 9.41\%. Furthermore, Figure \ref{resEconomyGraspxDFSxBFS} shows that the improved BC presents tighter gaps for almost the entire set of {\em Portifolio} instances than the original one.

In the third experiment, we investigate the behavior of the {\em UNGA} instances.
We notice that these instances are extremely easy to solve. No matter the number of vertices or the parameters used to compose the instance, both BC codes were always able to solve all of them in a few seconds and in the root of the branch-and-bound tree. 
So, we could not draw any conclusion from this class of instance.

In our last experiment, both methods were applied to each one of the 316 new {\em DMERN} instances~\cite{Figueiredo12MBS}. 
Table~\ref{resBC_NewDMERNInstances} shows the results for the instances remaining unsolved and the instances solved to optimality in more than one minute. The first three columns in this table give us information about the
instances: the Netlib instance name, the number of vertices and the number of
edges. The next three columns give the number of negative, positive and parallel edges, respectively.  
Similarly to the previous table, the next set of three columns gives us information about the solution obtained with the original BC code: the time, the percentage gap, and the total number of nodes in the branch-and-bound tree. The last three columns give the same data for the improved BC. From this set of instances, we can extract 25 instances not solved to
optimality by the original BC code with average gap of 11.42\% of unsolved instances, while the improved BC could not solve 21 instances but with a much tighter average gap of 4.85\%. 
One can notice that the implementation of new separation routines and a new branching rule used in the improved BC led to a better performance and a high number of evaluated nodes within the time limit.

\section{Final remarks}
\label{sec:Conclusion}

In this work, we proposed an improved branch-and-cut algorithm based on the integer programming formulation and the BC algorithm 
proposed in \cite{Figueiredo11}, together with a new branching rule based on the odd negative cycle inequalities and improved
 cutting plane routines and strategies. The instance classes reported in \cite{Figueiredo12MBS} were used to compare 
the performances of the improved BC and the original BC algorithm proposed in \cite{Figueiredo11}.
The results obtained by the new approach were superior to those
given by the previously existing branch-and-cut. The new method solved 431 out of 1445 instances within
1 hour of processing time, while the original algorithm managed to solve only 410 instances. 
Moreover, as we saw in Section~\ref{sec:Computational}, considering only the set of unsolved instances, the average gap obtained with 
the improved BC was smaller than the average gap obtained with the original BC from~\cite{Figueiredo11}.

\begin{sidewaystable}[Hp]
\footnotesize
\begin{center}
\begin{tabular}{r||rrrr||rrrr|rrr|rrr|rrr}
\hline
\multicolumn{1}{c||}{Instances} &
\multicolumn{4}{c||}{Time} &
\multicolumn{13}{c}{\%Gap} \\
\multicolumn{1}{c||}{} &
\multicolumn{4}{c||}{$|V|$} &
\multicolumn{4}{c|}{$|V|$} &
\multicolumn{3}{c|}{$d$} &
\multicolumn{3}{c|}{$|E^-|/|E^+|$} &
\multicolumn{3}{c}{$(|E^-\cap E^+|)/|E|$} \\
& \multicolumn{1}{c}{50}&
 \multicolumn{1}{c}{100}&
 \multicolumn{1}{c}{150}&
 \multicolumn{1}{c||}{200}&
 \multicolumn{1}{c}{50}&
 \multicolumn{1}{c}{100}&
 \multicolumn{1}{c}{150}&
 \multicolumn{1}{c|}{200}&
 \multicolumn{1}{c}{.25} &
 \multicolumn{1}{c}{.50} &
 \multicolumn{1}{c|}{.75} &
 \multicolumn{1}{c}{.50}  &
 \multicolumn{1}{c}{1}   &
 \multicolumn{1}{c|}{2}   &
 \multicolumn{1}{c}{.25} &
 \multicolumn{1}{c}{.50} &
 \multicolumn{1}{c}{.75} \\
\hline\hline
Group 1 &24.22(27)&2578.00(3)&  $-$     &  $-$  & 0	 & 37.05 & 104.55 &	153.42 & 75.48  & 88.03	& 82.83	& 75.84	& 86.01& 80.31&$-$&$-$&$-$\\	
        &10.63(27)&1728.33(9)&  $-$     &  $-$  & 0	 & 26.62 &	92.09 &	144.34 & 65.26  & 81.27 & 76.36 & 67.16 & 76.48& 74.27&$-$&$-$&$-$\\
\hline
Group 2 &2.41(27)&473.90(21)&1277.67(9)& $-$ & 0	 & 6.17	 &  49.08 &	111.83 & 33.48	& 56.28	& 65.78 &$-$&$-$&$-$& 68.69 & 42.22 & 21.35 \\
        &2.37(27)&323.33(21)&910.78(9)& $-$ & 0	 & 4.84  &	44.07 &	104.36 & 30.74  & 50.92 & 61.97 &$-$&$-$&$-$& 63.84 & 38.71 & 18.74 \\
\hline\hline
\end{tabular}
\end{center}
\caption{\footnotesize Results obtained on random instances in Group 1 ($E^-\cap E^+=\emptyset$) and in Group 2 ($E^-\cap E^+\neq\emptyset$).}
\label{resRandomInstBC}
\normalsize
\end{sidewaystable}

\begin{table}[Hp]
\scriptsize
\begin{center}
\begin{tabular}{rr|rrr|rrr}
\hline
\multicolumn{2}{c|}{Instance} &
\multicolumn{3}{c|}{Original BC} &
\multicolumn{3}{c}{Improved BC} \\
$|V|$	&	$t$	&		Time	&	\%Gap	&	Nodes	&		Time & \%Gap & Nodes  \\
\hline
330	&	0.300	&		25.00(2)	&	10.66	&	890.70	&		183.33(3)	&	4.56	&	933.50	\\
	&	0.325	&		295.25(8)	&	4.61	&	467.40	&		83.13(8)	&	2.82	&	431.60	\\
	&	0.350	&		13.00(10)	&	-	&	13.60	&		21.30(10)	&	-	&	34.80	\\
	&	0.375	&		1.50(10)	&	-	&	1.80	&		1.80(10)	&	-	&	2.70	\\
	&	0.400	&		1.00(10)	&	-	&	1.00	&		1.00(10)	&	-	&	1.00	\\
\hline
360	&	0.300	&		1145.67(3)	&	19.24	&	561.90	&		195.67(3)	&	6.48	&	581.20	\\
	&	0.325	&		170.75(4)	&	4.05	&	611.90	&		331.00(5)	&	2.39	&	914.20	\\
	&	0.350	&		161.10(10)	&	-	&	100.90	&		129.90(10)	&	-	&	135.50	\\
	&	0.375	&		3.10(10)	&	-	&	2.20	&		3.90(10)	&	-	&	4.40	\\
	&	0.400	&		1.10(10)	&	-	&	1.40	&		1.20(10)	&	-	&	1.50	\\
\hline
390	&	0.300	&		141.00(1)	&	29.52	&	498.80	&		650.50(2)	&	10.74	&	472.30	\\
	&	0.325	&		255.50(4)	&	17.15	&	461.80	&		101.25(4)	&	4.41	&	511.40	\\
	&	0.350	&		81.71(7)	&	2.40	&	372.80	&		29.14(7)	&	1.84	&	551.30	\\
	&	0.375	&		4.30(10)	&	-	&	2.40	&		5.20(10)	&	-	&	4.40	\\
	&	0.400	&		1.30(10)	&	-	&	1.10	&		1.40(10)	&	-	&	1.70	\\
\hline
420	&	0.300	&		-	&	30.56	&	401.70	&		-	&	15.86	&	395.70	\\
	&	0.325	&		1062.50(2)	&	13.63	&	432.30	&		1442.33(3)	&	8.24	&	548.30	\\
	&	0.350	&		176.14(7)	&	12.04	&	285.90	&		116.29(7)	&	3.98	&	322.60	\\
	&	0.375	&		192.10(10)	&	-	&	131.70	&		155.20(10)	&	-	&	201.10	\\
	&	0.400	&		7.40(10)	&	-	&	15.60	&		4.40(10)	&	-	&	11.50	\\
\hline
450	&	0.300	&		-	&	35.86	&	313.70	&		-	&	14.45	&	330.40	\\
	&	0.325	&		342.00(1)	&	14.75	&	360.40	&		124.00(1)	&	5.24	&	375.80	\\
	&	0.350	&		444.00(8)	&	2.40	&	241.70	&		390.89(9)	&	2.56	&	248.20	\\
	&	0.375	&		18.10(10)	&	-	&	8.40	&		24.00(10)	&	-	&	17.20	\\
	&	0.400	&		2.40(10)	&	-	&	1.30	&		2.70(10)	&	-	&	1.00	\\
\hline
480	&	0.300	&		2065.00(1)	&	42.69	&	243.60	&		740.00(1)	&	30.20	&	261.10	\\
	&	0.325	&		1746.33(2)	&	27.53	&	321.40	&		546.33(3)	&	13.66	&	298.10	\\
	&	0.350	&		385.20(5)	&	10.33	&	288.70	&		218.80(5)	&	3.43	&	318.80	\\
	&	0.375	&		43.22(9)	&	1.20	&	105.30	&		170.90(10)	&	-	&	83.40	\\
	&	0.400	&		23.90(10)	&	-	&	25.90	&		7.30(10)	&	-	&	7.00	\\
\hline
510	&	0.300	&		2809.00(1)	&	49.59	&	199.50	&		943.50(2)	&	33.17	&	182.60	\\
	&	0.325	&		392.00(2)	&	34.39	&	217.40	&		459.00(2)	&	19.92	&	244.70	\\
	&	0.350	&		47.00(3)	&	12.36	&	242.30	&		59.67(3)	&	3.70	&	315.70	\\
	&	0.375	&		101.29(7)	&	1.05	&	299.70	&		670.89(9)	&	0.53	&	563.90	\\
	&	0.400	&		6.60(10)	&	-	&	4.00	&		7.60(10)	&	-	&	4.40	\\
\hline
	&		&		(217)	&	17.91	&		&		(227)	&	9.41	&		\\
\hline
\end{tabular}
\end{center}
\caption{\footnotesize Results obtained on portfolio instances.}
\label{resBC_EconomyInstnaces}
\normalsize
\end{table}

\begin{sidewaystable}[Hp]
\scriptsize
\begin{center}
\begin{tabular}{lrrrrr|rrr|rrr}
\hline
\multicolumn{6}{c|}{Instance} &
\multicolumn{3}{c|}{Original BC} &
\multicolumn{3}{c}{Improved BC} \\
Name & $n$ & $m$ & $m-$ & $m+$ & $m-+$ & Time & \%Gap & Nodes & Time & \%Gap & Nodes	\\
\hline\hline
danoint	&	144	&	1456	&	497	&	903	&	56	&	289(1)	&	-	&	4349	&	164(1)	&	-	&	3951	\\
bienst1	&	184	&	2548	&	1981	&	567	&	0	&	360(1)	&	-	&	2523	&	2755(1)	&	-	&	39710	\\
stein45	&	331	&	10701	&	10701	&	0	&	0	&	2263(1)	&	-	&	651	&	-	&	4.03	&	508	\\
disctom	&	399	&	30000	&	30000	&	0	&	0	&	-	&	14.05	&	68	&	642(1)	&	-	&	16	\\
fc.60.20.1	&	414	&	1051	&	521	&	530	&	0	&	181(1)	&	-	&	399	&	172(1)	&	-	&	399	\\
air05	&	426	&	30257	&	30257	&	0	&	0	&	-	&	33.73	&	94	&	-	&	30.98	&	95	\\
neos17	&	486	&	117855	&	117370	&	0	&	485	&	38(1)	&	-	&	1	&	60(1)	&	-	&	1	\\
p100x588	&	688	&	1470	&	625	&	845	&	0	&	64(1)	&	-	&	71	&	62(1)	&	-	&	71	\\
air04	&	823	&	55592	&	55592	&	0	&	0	&	-	&	164.00	&	21	&	-	&	40.43	&	27	\\
r80x800	&	880	&	2000	&	1026	&	974	&	0	&	727(1)	&	-	&	223	&	699(1)	&	-	&	223	\\
nug08	&	912	&	13952	&	13952	&	0	&	0	&	75(1)	&	-	&	1	&	29(1)	&	-	&	1	\\
p50x864	&	914	&	1872	&	895	&	977	&	0	&	116(1)	&	-	&	53	&	113(1)	&	-	&	53	\\
dsbmip	&	1003	&	3733	&	2264	&	1383	&	86	&	70(1)	&	-	&	1	&	56(1)	&	-	&	1	\\
n5-3	&	1012	&	10750	&	5472	&	5278	&	0	&	66(1)	&	-	&	1	&	83(1)	&	-	&	1	\\
neos21	&	1085	&	37373	&	37373	&	0	&	0	&	-	&	274.67	&	24	&	783(1)	&	-	&	3	\\
neos23	&	1120	&	23387	&	22295	&	1092	&	0	&	109(1)	&	-	&	8	&	29(1)	&	-	&	2	\\
n4-3	&	1178	&	15341	&	7670	&	7671	&	0	&	139(1)	&	-	&	3	&	167(1)	&	-	&	1	\\
dano3mip	&	1227	&	46506	&	14948	&	31003	&	555	&	-	&	78.65	&	36	&	-	&	85.43	&	43	\\
n8-3	&	1300	&	11656	&	6258	&	5398	&	0	&	93(1)	&	-	&	1	&	119(1)	&	-	&	1	\\
roll3000	&	1300	&	60706	&	25022	&	31630	&	4054	&	693(1)	&	-	&	13	&	169(1)	&	-	&	2	\\
neos20	&	1320	&	14639	&	10788	&	3851	&	0	&	524(1)	&	-	&	75	&	106(1)	&	-	&	10	\\
p200x1188c	&	1388	&	2970	&	1228	&	1742	&	0	&	-	&	0.59	&	479	&	-	&	0.59	&	489	\\
p200x1188	&	1388	&	2970	&	1256	&	1714	&	0	&	-	&	0.63	&	494	&	-	&	0.63	&	519	\\
janos-us-ca--D-D-M-N-C-A-N-N	&	1643	&	11651	&	5491	&	6160	&	0	&	233(1)	&	-	&	1	&	213(1)	&	-	&	1	\\
pioro40--D-B-M-N-C-A-N-N	&	1649	&	10243	&	5777	&	4466	&	0	&	101(1)	&	-	&	1	&	126(1)	&	-	&	1	\\
n13-3	&	1661	&	14725	&	7579	&	7146	&	0	&	201(1)	&	-	&	1	&	215(1)	&	-	&	1	\\
n2-3	&	1752	&	14856	&	7935	&	6921	&	0	&	234(1)	&	-	&	1	&	259(1)	&	-	&	1	\\
qap10	&	1820	&	35200	&	35200	&	0	&	0	&	228(1)	&	-	&	1	&	424(1)	&	-	&	3	\\
ns1688347	&	1866	&	36800	&	24983	&	10195	&	1622	&	-	&	18.29	&	138	&	-	&	20.49	&	129	\\
ns25-pr3	&	1878	&	4333	&	1393	&	2940	&	0	&	112(1)	&	-	&	91	&	11(1)	&	-	&	7	\\
ns4-pr3	&	1878	&	4333	&	1393	&	2940	&	0	&	111(1)	&	-	&	91	&	10(1)	&	-	&	7	\\
ns60-pr3	&	1878	&	4333	&	1393	&	2940	&	0	&	111(1)	&	-	&	91	&	11(1)	&	-	&	7	\\
nu120-pr3	&	1878	&	4333	&	1393	&	2940	&	0	&	110(1)	&	-	&	91	&	10(1)	&	-	&	7	\\
nu25-pr3	&	1878	&	4333	&	1393	&	2940	&	0	&	110(1)	&	-	&	91	&	11(1)	&	-	&	7	\\
nu4-pr3	&	1878	&	4333	&	1393	&	2940	&	0	&	110(1)	&	-	&	91	&	10(1)	&	-	&	7	\\
nu60-pr3	&	1878	&	4333	&	1393	&	2940	&	0	&	110(1)	&	-	&	91	&	11(1)	&	-	&	7	\\
germany50--U-U-M-N-C-A-N-N	&	2088	&	10560	&	1143	&	2691	&	6726	&	13(1)	&	-	&	1	&	89(1)	&	-	&	1	\\
protfold	&	2112	&	89677	&	30219	&	58395	&	1063	&	-	&	53.07	&	3	&	-	&	53.40	&	4	\\
cap6000	&	2174	&	11167	&	10297	&	0	&	870	&	111(1)	&	-	&	1	&	110(1)	&	-	&	1	\\
n7-3	&	2278	&	24476	&	12220	&	12256	&	0	&	1431(1)	&	-	&	3	&	1184(1)	&	-	&	3	\\
n9-3	&	2280	&	33180	&	16280	&	16900	&	0	&	-	&	0.09	&	4	&	1321(1)	&	-	&	3	\\
acc-1	&	2286	&	44595	&	30912	&	13683	&	0	&	-	&	52.77	&	11	&	-	&	2.86	&	20	\\
n3-3	&	2303	&	38857	&	18602	&	20255	&	0	&	-	&	4.45	&	8	&	2821(1)	&	-	&	5	\\
zib54--D-B-E-N-C-A-N-N	&	2347	&	10025	&	6991	&	3034	&	0	&	236(1)	&	-	&	1	&	211(1)	&	-	&	1	\\
n12-3	&	2358	&	26496	&	12956	&	13540	&	0	&	1341(1)	&	-	&	1	&	1049(1)	&	-	&	1	\\
neos818918	&	2400	&	10130	&	6485	&	3195	&	450	&	819(1)	&	-	&	17	&	803(1)	&	-	&	17	\\
germany50--D-B-M-N-C-A-N-N	&	2438	&	12232	&	6325	&	5907	&	0	&	278(1)	&	-	&	1	&	260(1)	&	-	&	1	\\
acc-2	&	2520	&	60669	&	43842	&	16827	&	0	&	-	&	6.12	&	29	&	-	&	8.76	&	23	\\
ta2--U-U-M-N-C-A-N-N	&	2578	&	12312	&	2582	&	1834	&	7896	&	21(1)	&	-	&	1	&	173(1)	&	-	&	1	\\
n6-3	&	2686	&	31228	&	14664	&	16564	&	0	&	-	&	-	&	1	&	2753(1)	&	-	&	3	\\
berlin	&	2704	&	6630	&	2703	&	3927	&	0	&	-	&	0.94	&	16	&	-	&	0.94	&	17	\\
neos11	&	2706	&	47185	&	33685	&	13440	&	60	&	-	&	5.72	&	19	&	-	&	5.84	&	7	\\
ta2--D-B-M-N-C-A-N-N	&	2837	&	13457	&	9090	&	4367	&	0	&	380(1)	&	-	&	1	&	464(1)	&	-	&	1	\\
acc-6	&	3047	&	74184	&	55567	&	18571	&	46	&	-	&	11.09	&	14	&	-	&	11.09	&	10	\\
acc-5	&	3052	&	74312	&	54569	&	19697	&	46	&	-	&	14.30	&	11	&	-	&	13.84	&	11	\\
mkc	&	3127	&	6299	&	3503	&	2793	&	3	&	329(1)	&	-	&	1	&	338(1)	&	-	&	1	\\
mod011	&	3240	&	8186	&	8186	&	0	&	0	&	401(1)	&	-	&	1	&	431(1)	&	-	&	1	\\
acc-3	&	3249	&	72072	&	49812	&	22179	&	81	&	223(1)	&	-	&	1	&	225(1)	&	-	&	1	\\
acc-4	&	3285	&	75186	&	52301	&	22804	&	81	&	242(1)	&	-	&	1	&	241(1)	&	-	&	1	\\
brasil	&	3364	&	8265	&	3363	&	4902	&	0	&	-	&	0.85	&	9	&	-	&	0.85	&	9	\\
p500x2988c	&	3488	&	7470	&	3650	&	3820	&	0	&	-	&	4.59	&	68	&	-	&	4.52	&	70	\\
p500x2988	&	3488	&	7470	&	3064	&	4406	&	0	&	-	&	1.22	&	59	&	-	&	1.19	&	62	\\
rentacar	&	4294	&	16669	&	7916	&	8716	&	37	&	3043(1)	&	-	&	3	&	2380(1)	&	-	&	2	\\
neos1	&	4732	&	80870	&	41850	&	36380	&	2640	&	-	&	8.81	&	3	&	-	&	7.92	&	2	\\
seymour1	&	4794	&	604007	&	604007	&	0	&	0	&	-	&	14.42	&	0	&	-	&	15.25	&	0	\\
seymour	&	4794	&	604007	&	604007	&	0	&	0	&	-	&	14.42	&	0	&	-	&	15.25	&	0	\\
n370a	&	5150	&	15000	&	15000	&	0	&	0	&	1320(1)	&	-	&	1	&	1322(1)	&	-	&	1	\\
manna81	&	6480	&	72900	&	72900	&	0	&	0	&	439(1)	&	-	&	1	&	1173(1)	&	-	&	1	\\
neos12	&	8317	&	320726	&	302967	&	17549	&	210	&	-	&	10.38	&	0	&	-	&	10.38	&	0	\\
\hline
	&		&		&		&		&		&	413.75(43)	&	11.42	&	154.49	&	518.06(48)	&	4.85	&	675.26	\\
\hline
\end{tabular}
\end{center}
\caption{\footnotesize Results obtained on the new DMERN instances.}
\label{resBC_NewDMERNInstances}
\normalsize
\end{sidewaystable}




\bibliographystyle{plain}
\bibliography{BC_2.0}


\end{document}